\documentstyle[12pt]{article}
\begin{document}

Reply to the Comment made by Matthias Heger and Martin A. Suhm on:\\ ``Quantum Confinement in Hydrogen Bond"  
by Carlos da Silva dos Santos, Elso Drigo Filho and Regina Maria Ricotta, Int. J. Quantum Chem. 2015, 115, 765-770 \\
DOI: 10.1002/qua.24894\\

Recently we have presented a study that describes the use of an analytical technique to provide the shift on the frequency of the OH and NH groups present in a series of biological macromolecules. The shift in the vibrational frequency occurs because of the formation of hydrogen bonds between the group and a donor atom of the macromolecule, inducing quantum confinement, \cite{HBond}.  In order to describe such proposition we have employed the variational method associated with Supersymmetric Quantum Mechanics, SQM, which is essentially an algebraic method of quantum mechanics, to provide the energy spectra of the system before the hydrogen bond formation (system unconfined) and after the hydrogen bond formation (confined system).  The  main virtue of our method is  its simplicity.

The Schr$\ddot {o}$dinger equation was written with the Morse potential, notably a phenomenological potential with exact analytical solution, introduced in 1929 by Morse to treat molecular problems, \cite{Morse}.  The unconfined problem is therefore exact whereas the confined problem (with the hydrogen bond) has to have an approximation method to solve the Schr$\ddot {o}$dinger equation in a region surrounded by infinite walls, characterizing the confinement space where the hydrogen vibrates. The infinite walls are settled at the edges of the two atoms around the hydrogen, meaning that the hydrogen cannot penetrate inside them. The main point is the use of the cut off terms  $(y_{max}-y)(y-y_{min})$, which multiply the unconfined (exact) wave function to describe the confined wave function, where $(y_{max}-y_{min})$ is the region of space where the hydrogen vibrates. These two parameters, $y_{min}$ and $y_{max}$, are related to the covalent radius of the atom for which the hydrogen is chemically bound and the van der Waals radius complementary of hydrogen bond, respectively. The other parameters involved are the equilibrium distance, the energy dissociation and the reduced mass, which are characteristic of the Morse potential. 

The variational functions imply an approximation of what would be the actual functions for the studied problem and thus imply the assumption that the function would be close to the real one. The variational parameters used to minimize the energy eigenvalues must also reflect the correction of any distortions that alienate such a solution from the real solution of the problem. There is no evidence that this does not occur in this study. Rather, the convergence between experimental and theoretical values indicates that the suggested wave functions are appropriate for describing the desired system. 

It should be remarked that the Morse potential is not a first principles potential, but a potential constructed from the phenomenology. Seeking a conceptual reach the same potential is used for different environmental conditions where the hydrogen bonds appear. The quantitative results presented in the literature confirm this practice. There are no arguments or strong evidence at the moment to abandon this attitude in the incorporation of confinement to the description of hydrogen bonds in different media. Thus, it is understood that the use of experimental results for gases, as suggested by Heger and Suhm, can be used for further tests to the results presented, but do not invalidate the results obtained in \cite{HBond}.

We stress that the main objective of our work is to support the hypothesis that the confinement should be an important component for the description of hydrogen bonds and that the results, when the confinement is incorporated, describe the experimental data studied without exceeding the errors declared.

Despite the arguments raised, it is not clear that the quantum confinement hypothesis in the description of hydrogen bonds should be ruled out. Even though there is need for further investigation in some cases, particularly with respect to the conjecture that the confinement would necessarily lead to a decrease in the observed transition energy, as indicated in their comment, this conjecture, very strong, requires rigorous proof based on quantum calculation.

In the comment it seems there exists a confusion between a phenomenological description and the actual cause operating in the analyzed system, which leads to an effective distinction between an effective potential and a potential of first principles, respectively. Thus, the causes of hydrogen bond may be related to polarization and the weakening of the chemical bond as the authors emphasize and still be well described by the confined Morse potential. A good phenomenological potential should describe the phenomenon without necessarily explicitly include all causes. Take the Morse potential as an example; it has a similar shape to the Lennard-Jones potential, even though it does not explicitly display the dipole interaction term ($\propto 1/R^6$). Accordingly, the dipole term is embedded, but not explicitly when using the Morse potential to describe the system.

At last it should be mentioned newer results concerning the theoretical description of spectroscopic properties of DNA and RNA macromolecules, which corroborate all the arguments above, \cite{HBond-DNA}.

\end{document}